\documentclass[apj]{emulateapj}

\newcommand {\Lya}    {Ly$\alpha$}   
\newcommand {\Lyb}    {Ly$\beta$}    
\newcommand {\Lyg}    {Ly$\gamma$}
\newcommand {\Lyd}    {Ly$\delta$}
\newcommand {\HI}     {\ion{H}{1}}   

\newcommand {\OVI}    {\ion{O}{6}}   
\newcommand {\OVII}   {\ion{O}{7}}
\newcommand {\OVIII}  {\ion{O}{8}}
\newcommand {\CIII}   {\ion{C}{3}}   
\newcommand {\CIV}    {\ion{C}{4}}

\newcommand {\NV}     {\ion{N}{5}}

\newcommand {\SiIII}  {\ion{Si}{3}}

\newcommand {\kms}    {km~s$^{-1}$}
\newcommand {\NHI}    {$N_{\rm HI}$}

\newcommand {\hst}  {{\sl HST}}

\newcommand {\etal}  {et~al.}
\newcommand {\cd}    {cm$^{-2}$}

\slugcomment{Draft version of \today}
\shorttitle{Warm Gas in Spiral-Rich Group}
\shortauthors{Stocke \etal}

\begin{document}

\title{The Warm Circum-Galactic Medium: 10$^{5-6}$ K Gas Associated with a
Single Galaxy Halo or with an Entire Group of Galaxies?
\footnote{Based on observations with the NASA/ESA {\sl Hubble Space
Telescope}, obtained at the Space Telescope Science Institute, which
is operated by AURA, Inc., under NASA contract NAS 5-26555.}}

\author{John T. Stocke,
Brian A. Keeney,
Charles W. Danforth,
Benjamin D. Oppenheimer,
Cameron T. Pratt}
\affil{Center for Astrophysics and Space Astronomy, Department of
Astrophysical and Planetary Sciences, University of Colorado, 389 UCB,
Boulder, CO 80309, USA; john.stocke@colorado.edu}
\and
\author{Andreas A. Berlind}
\affil{Department of Physics and Astronomy, Vanderbilt University, 2310 
Vanderbilt Place, Nashville, TN 37235}

\shorttitle{Warm Spiral Group Gas?}
\shortauthors{Stocke et~al.}

\begin{abstract}
In preparation for a {\sl Hubble Space Telescope} (HST) observing
project using the Cosmic Origins Spectrograph (COS), the positions of
all AGN targets having high-$S/N$ far-UV G130M spectra were
cross-correlated with a large catalog of low-redshift galaxy groups
homogenously selected from the spectroscopic sample of the Sloan
Digital Sky Survey (SDSS).  Searching for targets behind only those
groups at $z=0.1$-0.2 (which places the \OVI\ doublet in the
wavelength region of peak COS sensitivity) we identified only one
potential $S/N=15$-20 target, FBQS\,1010$+$3003.  An \OVI-only
absorber was found in its G130M spectrum at $z=0.11326$, close to the
redshift of a foreground small group of luminous galaxies at
$z=0.11685$.  Because there is no associated \Lya\ absorption, any
characterization of this absorber is necessarily minimal; however, the
\OVI\ detection likely traces ``warm'' gas in collisional ionization
equilibrium at $T\approx3\times10^5$~K.  While this discovery is
consistent with being interface gas between cooler, photoionized
clouds and a hotter intra-group medium, it could also be warm,
interface gas associated with the circum-galactic medium (CGM) of the
single closest galaxy. In this case a detailed analysis of the galaxy
distribution (complete to $0.2\,L^*$) strongly favors the individual
galaxy association. This analysis highlights the necessity of both
high-$S/N>20$ COS data and a deep galaxy redshift survey of the region
in order to test more rigorously the association of \OVI-absorbing gas
with a galaxy group.  A Cycle 23 HST/COS program currently is
targeting 10 UV-bright AGN behind 12 low-redshift galaxy groups to
test the warm, group gas hypothesis.

\end{abstract}

\section{Introduction}
\label{intro}

A detailed knowledge of the Circumgalactic Medium (CGM; a.k.a. gaseous
galaxy halo) is necessary for any detailed understanding of galaxy
formation and evolution. Recent studies using the ultraviolet
spectrographs of the {\sl Hubble Space Telescope} (\hst) have proven
critical to recent advances in the field \citep{tripp98, penton04,
tumlinson11, prochaska11, stocke13, werk13, bordoloi14}, including the
recognition that the CGM likely contains a comparable number of
baryons as found in all the stars and gas in the disks of luminous
(L$\geq$ 0.1L$^*$) late-type galaxies \citep{tumlinson11, stocke13,
werk14, stern16, prochaska17, keeney17}. \citet {stocke13, stern16,
keeney17} estimate a CGM cool, photo-ionized gas mass somewhat less
than or comparable to the mass in stars and disk gas \citep[10-20\% baryon
fraction compared to $\sim$ 20\% in stars and disk gas; see Table~8
in][]{stocke13}, while \citet[][COS Halos project]{werk14,
prochaska17} obtain a limit and a value approximately a factor of 2
higher. While the COS-Halos results suggest that all the spiral galaxy
baryons may have been identified already, the other results suggest
that up to half the baryons are still ``missing''. The unaccounted gas
is likely in ``warm'' (10$^{5-6.5}$ K or hotter gas whose physical
conditions, extent and total baryonic content are not well-estimated
at the present time \citet {stocke14, werk16}.

By either mass estimate, the number of ``cool'' photoionized CGM
baryons is sufficient to explain the continuing high star formation
rate (SFR) in spiral galaxies \citep{binney87, chomiuk11} and the
detailed metallicity history of spiral galaxies like the Milky Way
\citep[e.g., the ``G dwarf problem'';][]{larson72, binney87,
chiappini01}, the number of detected CGM baryons is still less than
expected based both on detailed numerical simulations and on the
cosmic ratio of baryons to dark matter \citep[e.g.,][]{mcgaugh00,
klypin01}. Recent accountings by our group suggest that 30-50\% of
spiral galaxy baryons are still ``missing'' \citep{stocke13, keeney17}
in the sense that they have not been directly detected as yet \citep
{fukugita98, bregman07, shull12}.  Are these baryons still present in
the CGM of massive spirals or have they been ejected out into the
Inter-Galactic Medium (IGM) which is enriched in metals by this
process?

Sophisticated cosmological, numerical simulations
\citep[e.g.,][]{cen99, dave99} as well as scaling relations based on
X-ray observations of galaxy clusters and rich galaxy groups, suggest
that a hot intra-group gas should surround massive spirals in small
galaxy groups in the hard-to-detect temperature range of $10^{5-6.5}$
K \citep[see also][]{faerman16}.  This temperature range is too low to
provide thermal bremmstralung emission sufficient to be detected using
current X-ray telescopes \citep{mulchaey00}. Instead,
\citet{mulchaey96} suggested that this $T\approx10^6$ K spiral group
gas would be most easily detected using absorption-line spectroscopy
of background UV-bright sources.  Given the expected temperature range
for spiral group gas, the UV absorption doublet of \OVI\ 1032, 1038
\AA\ would be the most sensitive indicator.  However, the fraction of
oxygen which is in the quintuply-ionized state is small ($\leq1$\%) in
$T\geq10^6$ K gas with most of the oxygen being in more highly-ionized
states (\OVII\ and \OVIII), which are detectable in soft X-ray
absorption lines.  X-ray spectroscopy of low spectral resolution has
provided only possible low-S/N detections of group gas but potential
systematic noise in these very long integration {\sl Chandra} and {\sl
XMM-Newton} observations do not allow definitive detections or
non-detections of galaxy groups at this time \citep{nicastro10,
nicastro13, buote09}.  Potential Local Group detections are also
suspect because these can be attributed to the hot halo of the Milky
Way itself so that any \OVII\ absorptions detected at $z=0$ have
uncertain physical extent and origin \citep{bregman07}.  This leaves
broad, shallow \OVI\ and \Lya\ UV absorption lines as the best current
method for discovering hot gas in galaxy groups.  Statistical studies
\citep{stocke06, finn16} on the extent to which \OVI\ absorbers are
found away from galaxies suggest that metal enriched gas is spread
$\approx1$ Mpc from its source. This distance is about the diameter of
a small galaxy group. If this \OVI-absorbing gas fills the volume of
small galaxy groups at even modest filling factor, it could account
for the remainder of the missing baryons associated with spiral
galaxies like the Milky Way.

\citet[][Paper 1 hereafter]{savage14} have used high-S/N=20-50 Far-UV
(FUV) spectra obtained with the Cosmic Origins Spectrograph (COS)
\citep{green12} on HST to detect and analyze 54 \OVI\ absorbers at
low-$z$ including 14 systems which are estimated to be at temperatures
($T>10^5$ K) too hot to be photoionized gas. The remaining \OVI\
systems found in the Paper~1 study are either cooler than this,
suggesting photoionized gas, or have velocity misalignments that
preclude conclusive temperature analysis via this method (see Paper~1
for detailed analysis methodology). Two additional \OVI\ absorbers
have no associated \Lya\ arguing for very high temperatures where the
fraction of \HI\ is too low to be detectable. One of these two is the
initial HST/COS discovery based on the demonstration spectrum of
PKS\,0405$-$123 obtained by the COS Science Team just after
installation in Servicing Mission 5 \citep{savage10}.  Following
Paper~1 we will term this potential hot gas reservoir ``warm gas''
($T\sim10^{5-6.5}$~K), in contrast both to cool ($T\sim10^4$~K),
photo-ionized CGM gas commonly studied using COS FUV spectroscopy and
also the hot ($T\geq10^6$~K) intra-group and cluster gas detected with
X-ray telescopes.

The few broad, shallow absorptions seen in the ``warm'' \OVI$+$\Lya\
systems of Paper~1 are not obviously associated with individual
foreground galaxies, since these are found well outside the virial
radius of the nearest galaxy.  In \citet[][Paper~2
hereafter]{stocke14} we make the case that these warm absorbers are
associated with entire galaxy groups. If so a simple argument suggests
that these warm absorbers are very large and massive enough to account
for the remainder of the missing baryons in late-type galaxies;
viz. using the warm absorber line density of $d{\cal N}/dz\approx 4$
per unit redshift (Paper~2) in conjunction with the local space
density of galaxy groups \citep{berlind06, berlind08, pisani03}
requires that these absorbers have appproximate radii of 1 Mpc at high
covering factor (larger still if patchy) and therefore are quite
massive ($\sim 10^{12} M_\odot$; Paper~2). This is comparable to the
gas mass in E-dominated groups and poor clusters \citep{mulchaey00}
and is also the amount needed to bring spiral groups up to the cosmic
mean baryon-to-dark matter ratio \citep[see also the recent paper
by][]{faerman16}. The large uncertainty in the mass estimate above due
to the largely unknown extent and filling factor of the warm,
intra-group gas means that the baryon pecentage remaining within the
group's extent is still quite uncertain.

Current numerical simulations differ by factors of 3-5 on the
percentage of baryons retained by group-size halos in this critical
mass range ($\log (M/M_\odot)=13.0-14.0$).  COSMO-Owls
\citep{lebrun14}, EAGLE \citep{schaye15} and ILLUSTRIS \citep{genel14}
find divergent values in this halo mass range. But if all these
baryons are retained within the group, then these spiral galaxy groups
are close to being ``closed boxes'', which has important consequences
for galactic evolution. For example, the metallicity and
star-formation rate histories then must be reconciled within
individual galaxy groups and can be quite different from
group-to-group depending upon the group luminosity function and
details of the star-formation history. Given these important
consequences, the hypothesis of a massive warm gaseous reservoir in
spiral galaxy groups needs to be verified.

With these questions in-mind we have initiated an HST/COS program in
Cycle 23 to obtain $S/N \geq 20$ FUV spectra of ten bright AGN targets
whose sightlines penetrate galaxy groups selected homogenously from a
large new, low-$z$ group catalog that represents an extension of
\citet[][Paper 3 hereafter]{berlind06}. In preparation for this program
we searched the HST/COS archive for serendipitous detections of broad,
shallow \OVI\ associated with $z=0.1$-0.2 galaxy groups from this
catalog.  Despite having thousands of groups and hundreds of QSO
targets to use for this search, only six targets previously observed
by COS lie behind a Berlind group in the redshift range which places
any potential \OVI\ absorption in the wavelength band of peak COS
sensitivity.  All but one of these six have FUV spectra with S/N too
low for detecting broad, shallow warm absorbers similar to those
studied in Papers~1 and 2.  However, FBQS\,1010$+$3003 has
medium-quality archival COS data ($S/N=15$ at \OVI; and $S/N=20$ at
\Lya\, somewhat lesser in quality to what we proposed for in Cycle
23).  Here we report the discovery in the FUV spectrum of the bright
moderate redshift ($z=0.2564$) AGN FBQS\,1010$+$3003 of a broad
($b\approx 50$ \kms), shallow and symmetrical \OVI\ 1032\AA\ detection
at the redshift ($z=0.11326$) of a spiral-rich group (Berlind \#49980
at $z=0.11455$)\footnote{We note that this group and its ID are
associated with the new group catalog that we have constructed for
this paper and does not correspond to any previously published group
from Paper 3.}.

In Section 2 we briefly describe the selection process which allowed
this discovery.  Section 3 presents the COS spectra of \OVI\ and
\Lya\ at the group redshift. Section 4 describes the deep galaxy
redshift survey in this region which resulted in 30 galaxies at
$L\geq0.2 L^*$ which are potential group members.  The membership of
this galaxy group is also described in this Section.  A brief
discussion of this result and a summary of our Conclusions are
presented in Section 5.


\section{Sample Search and Selection}

To select targets we used a large new catalog of galaxy groups
($>7500$) that was constructed in a similar manner as in Paper~3, but
was based on a slightly higher redshift sample ($z=0.1$-0.2) from the
Sloan Digital Sky Survey (SDSS) Data Release 7 data
\citep{abazajian09}.  Restricting the group redshifts to $z=0.1-0.2$:
(1) allows good SDSS galaxy group membership selection and
characterization (estimated total luminosities, sizes and velocity
dispersions); (2) facilitates excellent follow-up observations using
multi-object galaxy spectroscopy (MOS) on moderate aperture telescopes
(already in progress); and (3) allows both the \OVI\ doublet and \Lya\
(as well as \Lyb, \SiIII, and other lines of potential interest) to be
observed with COS in the highest throughput G130M mode in a single
visit.  While there is an abundance of $z<0.1$ groups in the
\citet{berlind08} catalog and in other catalogs, the \OVI\ doublet is
not within the COS passband at those low redshifts.  Much of the
preferred redshift range for \OVI\ is blueward of the \Lya\ rest
wavelength, making an \OVI\ identification more secure even with the
detection of a single line in the \OVI\ doublet.

We cross-correlated the groups catalog with a list of bright
($V<17.5$) background AGN targets. For investigation of a sample of
sightlines through groups, we required that the AGN sightline
intersect these groups at a range of impact parameters
$0.25<(\rho/R_{grp})<1.5$, where $R_{grp}$ is the estimated virial
radius of the group based on the estimated virial mass, $M_{grp}$. The
virial mass was assigned by matching the abundance of groups of a
given total $r$-band luminosity to the theoretical abundance of dark
matter halos of a given mass, as derived from a standard concordance
cosmology halo mass function. For our Cycle 23 HST/COS GO program we
rejected sightlines which passed within 1.5 virial radii of an
individual group galaxy to make sure we were observing group gas, not
gas associated with individual galaxies \citep[see
e.g.,][]{prochaska11, stocke13}.  This impact parameter is large
enough that association of an absorber with a single galaxy is
problematical \citep[see Paper~2 and][] {keeney17}.  But this proved
too restrictive a criterion given the limited number of high-S/N
spectra in the current COS archive \citep{danforth16}.

A high-S/N ($>20$ per resolution element at the predicted wavelength
of \OVI\ 1032\AA) is essential for this program. The symmetrical and
aligned \OVI\ and \Lya\ collisionally ionized lines at $T\geq10^5$~K
are distinct from the large majority of \OVI$+$\Lya\ absorption
systems in the Paper~1 sample \citep[and those found in all other
samples at lower S/N; e.g.,][]{tumlinson11, borthakur13, bordoloi14}
since only 14 out of 54 \OVI\ systems from Paper~1 are unambiguously
warm gas. Broad, shallow and symmetrical \OVI\ absorbers without
associated \Lya\ are due unambiguously to warm, collisionally-ionized
gas \citep[see examples in][and Paper~1]{savage10}.

Despite our estimate based on Paper~2 that $S/N>20$ is required for a
warm gas detection, we cross correlated the Berlind groups with $z
=0.1-0.2$ with all available HST/COS G130M AGN spectra. Additionally,
we discarded the restriction that no individual galaxy can be within
1.5 virial radii of the sightlines.  Six matches were found within the
bounds of the group virial radius but only FBQS\,1010$+$3003 possessed
a COS G130M spectrum of anywhere near the quality required for a
detection like those in Papers~1 and 2. The remaining five targets
showed no sign of an \OVI\ absorption near the group redshift;
however, an absorption comparably strong to the FBQS\,1010$+$3003
detection would {\it not have been detected} in the other five low-S/N
spectra.

Expanding the above procedure to include bright targets not yet
observed by HST/COS found one dozen potential candidates, from which
we chose ten for observation in Cycle~23.  Between the time of our
initial search and the HST proposal submission, one of these ten
(CSO\,1022) was observed in the Cycle~22 HST Guest Observer program
\#13444 (PI: B. Wakker).  That observation will be reported along with
the results of our on-going Cycle~23 program at a later time.


\section{The \OVI-only system in FBQS\,1010$+$3003}

\begin{figure*}
   \epsscale{.9}\plotone{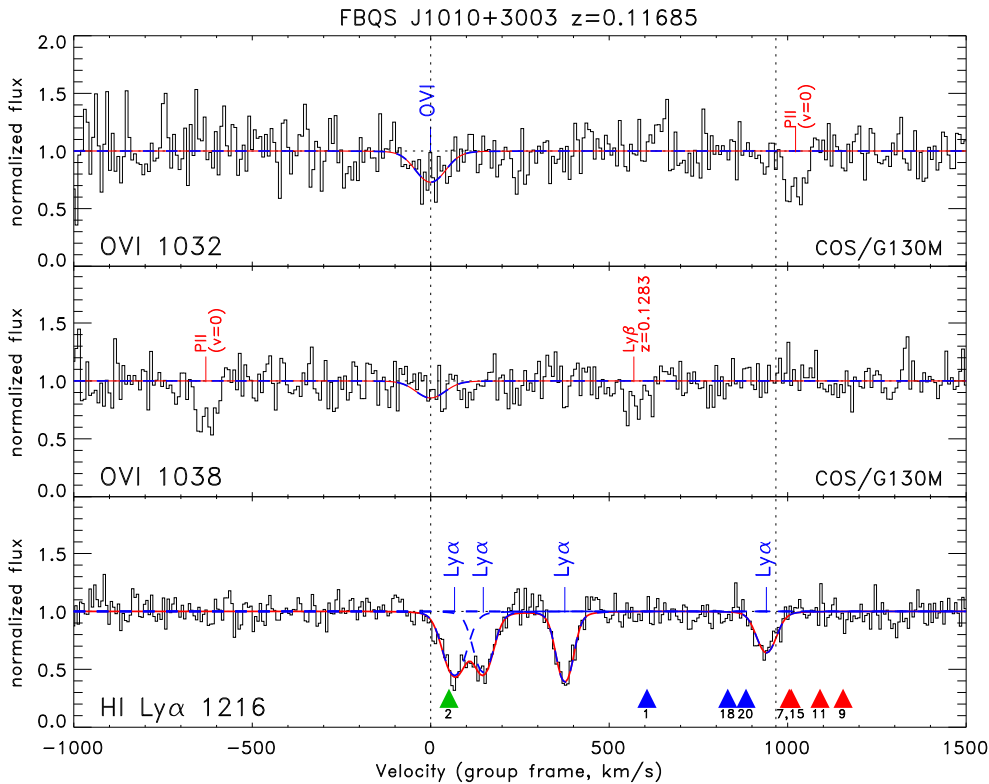} 
   \caption{The normalized COS spectrum of FBQS\,1010$+$3003
   corresponding to \OVI\ and \Lya\ absorption at $z\approx0.115$. The
   vertical dotted lines indicates the fiducial velocity of the \OVI\
   detection ($z=0.11326$, left) and the revised redshift of Berlind
   group \#49980 ($z=0.11685$, right).  Weak, broad \OVI\ is seen in
   the stronger line of the doublet (top panel) and is consistent with
   an upper limit in the weaker doublet line (middle panel).  Four
   \Lya\ absorption components are seen (bottom panel), none of which
   align with the \OVI\ line.  The offset to the closest component is
   50 \kms\ several times larger than either the COS resolution
   element or the uncertainties typically found in the COS wavelength
   scale.  Red lines are the absorption model while dashed blue lines
   show each fitted Voigt component.  The bottom panel shows triangles
   corresponding to the redshift of galaxies associated with the
   group.  Red and blue triangles indicate group members at higher or
   lower velocities, respectively of the group centroid (see text for
   discussion).  The green triangle at $v=50$ \kms\ indicates the
   galaxy \#2, the closest galaxy to the sightline and which is {\bf
   not} a group member.}
\end{figure*}

Figure 1 shows the wavelength regions of the normalized
FBQS\,1010$+$3003 COS/G130M spectrum at the \OVI\ doublet (top and
middle) and the \Lya\ region (bottom).  Each spectral region is
presented in velocity space along the x-axis with the origin at the
redshift of the \OVI\ absorber ($z=0.11326$).  The FBQS\,1010$+$3003
COS spectrum was obtained by the COS Science Team as part of its
Guaranteed Time Observation (GTO) program and reduced and analyzed as
described in \citet{danforth16}.  It has medium-quality with $S/N=15$
per resolution element at \OVI\ and 20 at \Lya\, somewhat lesser in
quality to what was predicted to be required for detection of typical
warm absorbers based on previous GTO spectra (Paper~2).
Table~\ref{tab:linefits} lists the best-fit Voigt profile particulars
for each line.

\begin{deluxetable*}{lcclllc}
   \tabletypesize{\footnotesize}
   \tablecolumns{7}
   \tablewidth{0pt}
   \tablecaption{Absorption Line Fit Parameters}
   \tablehead{
            \colhead{Transition}          &
            \colhead{Wavelength}           &
            \colhead{$z_{abs}$}          &
            \colhead{$W_r$} &
            \colhead{$b$} &
            \colhead{$\log N$} &
            \colhead{SL}              \\
            \colhead{}          &
            \colhead{(\AA)}           &
            \colhead{}          &
            \colhead{(m\AA)} &
            \colhead{(\kms)} &
            \colhead{(\cd)}  &
            \colhead{($\sigma$)}
             }
\startdata
 O\,VI 1032 & 1148.80 & 0.113258 &$102\pm57 $&$ 52\pm16$&$14.0\pm0.2  $ &  3.8\\
 \Lya\ 1215 & 1353.64 & 0.113508 &$195\pm31 $&$ 38\pm6 $&$13.69\pm0.07$ & 12.5\\
 \Lya\ 1215 & 1354.00 & 0.113807 &$158\pm18 $&$ 32\pm6 $&$13.60\pm0.06$ & 10.9\\
 \Lya\ 1215 & 1355.03 & 0.114655 &$162\pm11 $&$ 26\pm3 $&$13.65\pm0.04$ & 12.3\\
 \Lya\ 1215 & 1357.58 & 0.116752 &$105\pm14 $&$ 33\pm6 $&$13.37\pm0.06$ &  7.8\\
\enddata
   \label{tab:linefits}
\end{deluxetable*}

A broad, shallow absorption feature detected at $\sim4\sigma$ at
1148.80 \AA\ is identified as \OVI\ 1032\AA\ at $z=0.113258$ with
$\log N_{OVI}=14.0\pm0.2$ and $b_{OVI}=52\pm16$ \kms.  Ideally, this
identity would be confirmed with a corresponding feature in the weaker
line of the doublet at 1155.14\AA, however, the expected \OVI\
1037\AA\ feature would be of low significance ($\la2\sigma$), but its
marginal presence in the middle panel of Figure~1 is consistent with
its predicted strength, width, and wavelength location based on the
1032\AA\ detection (red and blue-dashed lines in Figure~1).
The correlated pixel (fixed-pattern) noise characteristics of the COS
detectors have yet to be quantified which makes it impossible to
calculate a formal reduced $\chi^2$ value for the 1037 \AA\
non-detection.  However, we can calculate a relative goodness-of-fit
for different \OVI\ 1037 absorber scenarios.  The 1037 \AA\ line
profile predicted by the 1032 \AA\ fit gives a $\chi^2$ value
identical to that of a flat continuum (no absorber) as well as that of
an OVI absorber of one third the strength as seen in 1032.

\Lya\ (bottom panel of Figure~2) shows several, narrower, photoionized 
CGM components most likely associated with individual group galaxies,
but there is no \Lya\ absorber that can be associated specifically
with the \OVI\ detection.

Identifying an absorption feature based on a single line is risky and
we must acknowledge that it is possible that the 1148.8\AA\ feature is
{\em not} \OVI.  However, in this case, the list of alternate
identifications for this significant absorption feature are extremely
meager.  The feature's location blueward of 1216\AA\ rules out a weak
\Lya\ forest line \citep[the default identification of weak lines
in][]{danforth16}.  The modest redshift of FBQS\,J1010$+$3003 ($z_{\rm
AGN}=0.2558$) means that the only other \HI\ Lyman transitions which
could possibly occur at 1148.8\AA\ (\Lyb\ $z=0.12000$, \Lyg\
$z=0.18124$, \Lyd\ $z=0.20960$, etc.) do not show corresponding
absorption in stronger Lyman lines at redward portions of the
spectrum.  \OVI\ is the most commonly-seen metal species in AGN sight
lines, but none of the other lines commonly seen in extensive surveys
\citep[\CIV, \SiIII, \CIII][]{danforth16} are consistent with the
redshift range and/or the presence of other absorption.  There are no
Galactic absorption features nor any known COS instrumental features
that fall near 1148.8\AA.  Similarly, the feature is far too narrow to
be an intrinsic feature of the AGN continuum itself.  \OVI\ 1032\AA\
is the only realistic identification.

There are no other plausible identifications for this absorption line
since it is found blueward of the \Lya\ rest wavelength.  Metal line
locations associated with \Lya\ absorption systems at other redshifts
and at the systemic redshift of FBQS\,1010$+$3003 were checked and
there are no reasonable possibilities for this absorption other than
\OVI.  To be conclusqively convincing and to determine a robust
$b$-value would require both lines in the doublet to be detected
independently.  However, the near wavelength coincidence between the
\OVI\ and the several \Lya\ absorbers (likely CGM absorption from
individual group galaxies; see Figure~1), adds credence to this
proposed identification.  Clearly, higher S/N data and detections
which include an associated broad \Lya\ line will be required to be
more definitive in other cases.

This \OVI\ absorber is similar to the case described in detail in
\citet{savage10} found in the spectrum of PKS\,0405$-$123 in which a
broad, symmetrical and shallow \OVI-only system was discovered.  As in
that paper we can use the observed \OVI\ line width and the limit on
the presence of \Lya\ to roughly bound the temperature and/or
metallicity of the gas producing the \OVI.

Since the total line width is the quadratic sum of thermal and
non-thermal motions: $b_{tot}^2=b_{therm}^2+b_{non-thermal}^2$, the
measured value for this \OVI\ 1032\AA\ line limits the temperature to
$T<2.6\times10^6$~K (i.e., assuming $b_{nt}=0$).  However, there is
considerable uncertainty in the line width due to the finite quality
of the data.  Taking the full $\pm1\sigma$ of the fitted $b$-value
($b_{OVI}=52\pm16$ \kms) under the $b_{nt}=0$ assumption gives a range
of upper limits to the temperature of $T<(1.3-4.5)\times10^6$~K.

The absence of obvious irregularities in the line profile of the \OVI\
1032\AA\ absorption suggests (but does not prove) that the line width
is largely thermal, in which case the suggested temperature is
$>10^5$~K.  The absence of corresponding low-ionization metal ions
often seen in IGM systems (e.g., \SiIII) also provides a lower limit
on the gas temperature of $\geq 2\times 10^5$~K, signbificantly hotter
than is typical of photoionization equilibrium (PIE) making
collisional ionization equilibrium (CIE) far more likely in this case
as well.  Other highly-ionized metal ions (e.g., \CIV, \NV) are seen
in some other ``warm'' absorbers, but the relatively weak \OVI\
absorption and the relative abundances of C and N compared to O makes
them undetectable in data of this modest quality.  We set $3\sigma$
non-detections for both \CIV\ and \NV\ of $\log N\la13$.  A complete
physical discussion showing how an \OVI\ detection and a \Lya\
non-detection lead to CIE and temperatures in excess of $10^5$~K for
the few \OVI-only systems we have discovered \citep[][and in
Paper~2]{savage10} can be found in \citet{savage10}.

\begin{figure*}
\epsscale{.95}\plotone{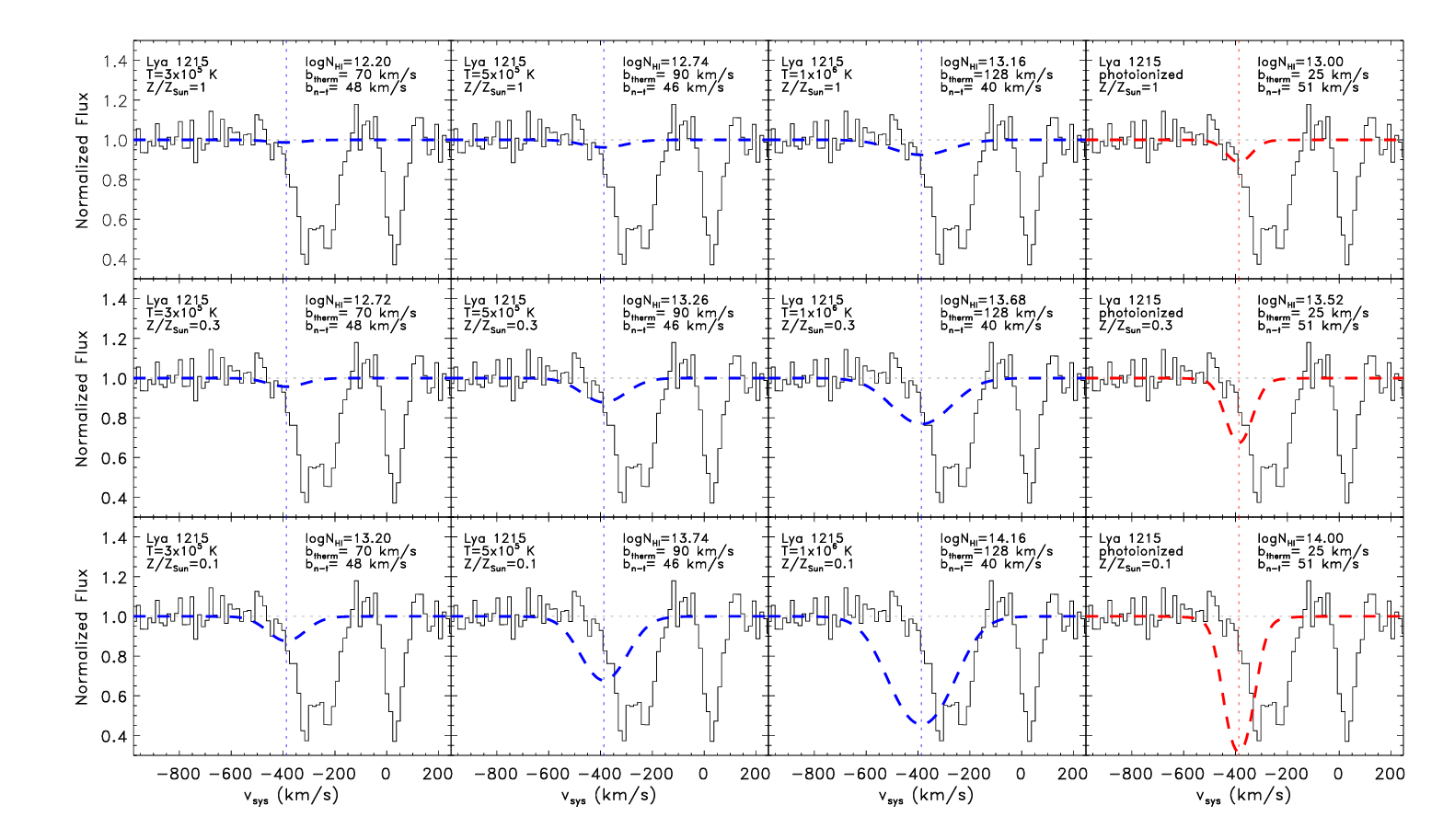}
   \caption{\Lya\ line profiles implied by the observed O\,VI profile
   fit ($N_{\rm OVI}=14.0\pm0.2$, $b_{\rm OVI,total}=52\pm16$ \kms)
   and a grid of assumed temperatures (left-right: $T=3\times10^5$,
   $5\times10^5$, $10^6$K) and metallicities (top-bottom: $Z=Z_\sun$,
   $0.3~Z_\sun$, $0.1~Z_\sun$).  Implied \NHI\ and thermal and
   non-thermal $b$ values are listed in each case.  The three strong
   absorption features are narrow \Lya\ components in Figure~1.  Note
   that the bluest \Lya\ line is misaligned with the O\,VI absorption
   by $\sim100$ \kms.  We qualitatively rule out low metallicities and
   high temperatures by this method.  The fourth column of models
   assumes a warm ($T=4\times10^4$ K) in which O\,VI is produced
   through photoionization.  We can rule this case out as well except
   in the case of solar or higher metallicity.}
\end{figure*}

Figure~2 displays a grid of synthetic \Lya\ spectra created assuming
the observed \OVI\ column density and line width, CIE and the
temperature and metallicity values displayed in each box.  While the
observational constraints are scant, temperatures near $3\times10^5$~K
and higher metallicities ($[Z]\geq -0.5$) are preferred given the
identification of the one observed detection as \OVI.

It is possible that the \OVI\ absorber is photoionized at a
temperature more typical of the \Lya\ forest rather than WHIM gas.
CLOUDY simulations \citep[e.g., those used in ][]{keeney17} show that
the ratio of ionization fractions of \OVI\ and \HI\ ($f_{HI}/f_{OVI}$)
remain fairly constant over a wide range of ionization parameters at
$T<10^5$~K.  Adopting the observed $N_{\rm OVI}$, we infer a
characteristic \NHI\ expected for photoionized gas of
$N_{HI}\sim10^{14}/Z_{0.1}$ for a canonical $Z/Z_{\odot}=0.1$.  The
observed \OVI\ profile must be broadened almost entirely by
non-thermal motions for gas at these low temperatures, so this
non-thermal broadening would be present in the inferred \Lya\ profiles
as well.  We show these predicted \Lya\ profiles under the
photoionized scenario and a temperature of $4\times10^4$~K in the
fourth column of Figure~2 (red dashed lines).  We can rule out
photoionization except in the case of solar metallicity or higher.


\section{The Galaxy Group Berlind \#49980}

\begin{figure}
   \epsscale{1.2}\plotone{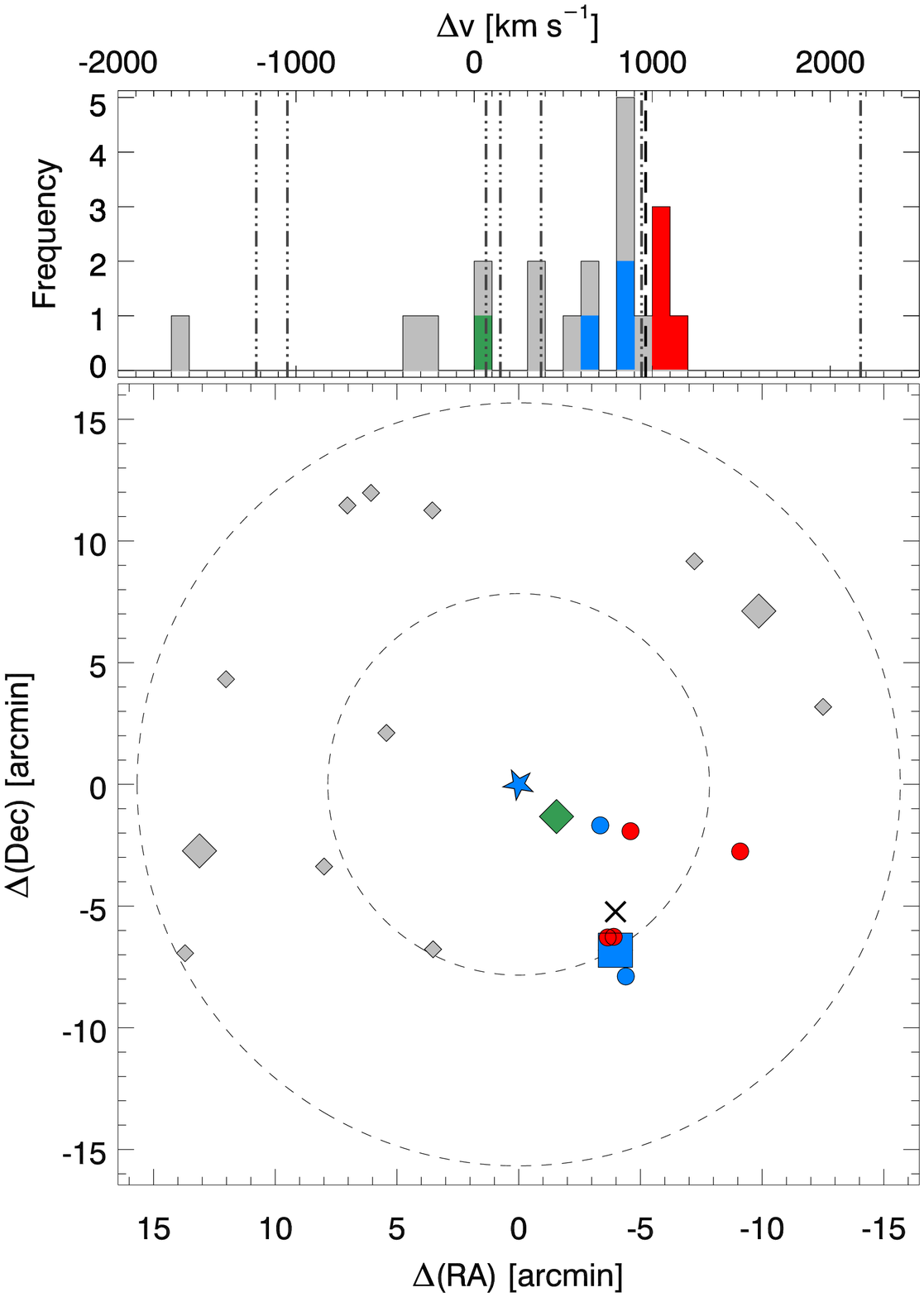}
   \caption{The galaxy group near the FBQS\,J1010$+$3003 sight line at
   $z_{\rm grp}=0.11685$. The top panel is a histogram of galaxy
   radial velocities relative to the \OVI\ absorber at $z_{\rm
   abs}=0.11326$.  The colored bins are the seven group members
   identified by the less conservative of two friends-of-friends
   algorithms (see text). The blue and red bins are group galaxies
   blueshifted and redshifted relative to the group centroid. The gray
   bins are the galaxies excluded by the group finding algorithm, and
   the green bin is the galaxy located closest to the QSO sight
   line. The dashed vertical line is the group velocity centroid and
   the dot-dashed vertical lines are absorption-line velocities of
   \HI\ \Lya\ in this velocity window. The bottom panel shows the
   spatial distribution of galaxies on the sky with respect to the QSO
   sight line, which is indicated by the large star at the origin. The
   dashed circles show impact parameters in 1~Mpc increments
   increasing outwards. The positions of group galaxies are marked
   with squares for redshift data from the SDSS or circles for
   redshift data from our own WIYN/HYDRA observations.  The positions
   of galaxies that are not group members are marked with
   diamonds. The symbol size indicates galaxy luminosity, with the
   largest symbols representing galaxies with $L>L^*$.  As in the
   histogram at top, blue and red symbols mark group members with
   $z_{\rm gal}<z_{\rm grp}$ and $z_{\rm gal}>z_{\rm grp}$,
   respectively, and the green symbol marks the location of the galaxy
   closest to the QSO sight line.  The large ``X'' is the group
   centroid on the plane of the sky.}
\end{figure}

\begin{deluxetable*}{cllccccr}

\tabletypesize{\scriptsize}

\tablecolumns{8}
\tablewidth{0pt}

\tablecaption{All Galaxies Near FBQS~J1010+3003 ($z_{\rm abs} = 0.11326$)}
\label{tab:gal_all}

\tablehead{ \colhead{ID} & 
            \colhead{Name} & 
            \colhead{Source} & 
            \colhead{$z_{\rm gal}$} & 
            \colhead{$L_{\rm gal}$} & 
            \colhead{$\rho$} & 
            \colhead{$\rho/R_{\rm vir}$} & 
            \colhead{$\Delta v$} \\ 
            & 
            & 
            & 
            & 
            \colhead{($L^*$)} & 
            \colhead{(kpc)} & 
            & 
            \colhead{(\kms)} } 

\startdata
 1 &  SDSS~J100942.33+295632.8 &  SDSS & 0.11551 & 2.158 &  979 &  4.19 & $  607$ \\
 2 &  SDSS~J100953.50+300202.2 &  SDSS & 0.11345 & 1.980 &  254 &  1.11 & $   51$ \\
 3 &  SDSS~J101101.29+300037.7 &  SDSS & 0.11206 & 1.886 & 1664 &  7.38 & $ -324$ \\
 4 &  SDSS~J100915.11+301028.8 &  SDSS & 0.11470 & 1.492 & 1510 &  7.29 & $  386$ \\
 5 &  SDSS~J101056.25+300740.7 & HYDRA & 0.11472 & 0.967 & 1586 &  8.80 & $  393$ \\
 6 &  SDSS~J101033.20+301449.2 &  SDSS & 0.11654 & 0.737 & 1669 & 10.15 & $  884$ \\
 7 &  SDSS~J100942.65+295705.8 & HYDRA & 0.11701 & 0.530 &  917 &  6.26 & $ 1009$ \\
 8 &  SDSS~J101025.80+300528.4 & HYDRA & 0.11350 & 0.491 &  724 &  5.06 & $   64$ \\
 9 &  SDSS~J100943.80+295704.2 & HYDRA & 0.11755 & 0.477 &  903 &  6.35 & $ 1155$ \\
10 &  SDSS~J101017.08+301437.0 & HYDRA & 0.11642 & 0.426 & 1466 & 10.72 & $  850$ \\
11 &  SDSS~J100939.46+300126.2 & HYDRA & 0.11731 & 0.419 &  619 &  4.56 & $ 1090$ \\
12 &  SDSS~J101016.94+295635.0 & HYDRA & 0.10727 & 0.404 &  948 &  7.04 & $-1613$ \\
13 &  SDSS~J101104.02+295625.4 & HYDRA & 0.11515 & 0.395 & 1908 & 14.28 & $  508$ \\
14 &  SDSS~J100927.32+301231.5 & HYDRA & 0.11635 & 0.255 & 1449 & 12.53 & $  832$ \\
15 &  SDSS~J100918.64+300036.2 & HYDRA & 0.11699 & 0.249 & 1181 & 10.31 & $ 1004$ \\
16 &  SDSS~J101037.63+295959.0 & HYDRA & 0.11219 & 0.234 & 1078 &  9.61 & $ -288$ \\
17 &  SDSS~J100902.90+300632.6 & HYDRA & 0.11578 & 0.232 & 1602 & 14.28 & $  678$ \\
18 &  SDSS~J100945.21+300140.5 & HYDRA & 0.11635 & 0.217 &  466 &  4.25 & $  832$ \\
19 &  SDSS~J101028.73+301520.0 & HYDRA & 0.11695 & 0.213 & 1667 & 15.34 & $  993$ \\
20 &  SDSS~J100940.35+295528.3 & HYDRA & 0.11654 & 0.212 & 1122 & 10.33 & $  883$
\enddata

\end{deluxetable*}

\begin{deluxetable*}{cllcccc}

\tabletypesize{\scriptsize}

\tablecolumns{7}
\tablewidth{0pt}

\tablecaption{Galaxy Group Members ($z_{\rm grp} = 0.11685$)}
\label{tab:gal_group}

\tablehead{ \colhead{ID} & 
            \colhead{Name} & 
            \colhead{$z_{\rm gal}$} & 
            \colhead{$L_{\rm gal}$} & 
            \colhead{$\rho$} & 
            \colhead{$R_{\rm sub}$} & 
            \colhead{$\log{M_*}$} \\ 
            & 
            & 
            & 
            \colhead{($L^*$)} & 
            \colhead{(kpc)} & 
            \colhead{(kpc)} &
	    \colhead{($M_\odot$)} } 

\startdata
18 &  SDSS~J100945.21+300140.5 & 0.11635 & 0.217 &  466 & 110  & 10.29 \\
11 &  SDSS~J100939.46+300126.2 & 0.11731 & 0.419 &  619 & 136  & 10.66 \\
 9 &  SDSS~J100943.80+295704.2 & 0.11755 & 0.477 &  903 & 142  & 10.02 \\
 7 &  SDSS~J100942.65+295705.8 & 0.11701 & 0.530 &  917 & 147  & 10.77 \\
 1 &  SDSS~J100942.33+295632.8 & 0.11551 & 2.158 &  979 & 234  & 11.41 \\
20 &  SDSS~J100940.35+295528.3 & 0.11654 & 0.212 & 1122 & 109  & 10.43 \\
15 &  SDSS~J100918.64+300036.2 & 0.11699 & 0.249 & 1181 & 115  & 10.03
\enddata

\end{deluxetable*}


Originally, the group Berlind \#49880 consisted of three luminous
galaxies culled from a short list of five galaxies from this sky area
($\pm20$ arcminutes in each coordinate) and redshift interval
($\pm1000$ \kms) which are in the SDSS spectroscopic sample.  These
three galaxies have a mean redshift of $z=0.114550$, a total estimated
halo mass (in solar masses) of $\log M_{grp}=13.53$ and an estimated
$R_{grp}=0.96$~Mpc.  The FBQS\,1010 sightline lies at $0.94\,R_{grp}$
from the group centroid.  The second-brightest galaxy in the group is
a $2\,L^*$ SB spiral with weak H$\alpha$ and [\ion{N}{2}] 6584\AA\
emission that is 254 kpc ($1.1\,R_{vir}$) from the sightline.  The
absorber/galaxy velocity difference is only 51~\kms\ making an
absorber association with this one galaxy entirely plausible.  All
other group galaxies are farther away from the sightline and at a much
larger number of virial radii given that the three other potential
group galaxies at an impact parameter $\rho<0.75$~Mpc are much less
luminous.  The two SDSS galaxies not selected for group membership
fall just below the luminosity limit of the $z=0.1$-0.2 volume-limited
sample used to identify groups.  It is thus useful to look at this
region again in the light of deeper, newly obtained redshift
information.  The five SDSS galaxies are listed in Table~2 in
decreasing luminosity (ID = 1-4 and \#6); the first three (ID = 1-3)
are the group members initially identified.

In order to characterize the galaxy group associated with this
absorber, multi-object spectroscopy (MOS) in the field of
FBQS\,1010$+$3003 was obtained using the HYDRA spectrograph on the
Wisconsin-Indiana-Yale-NOAO (WIYN) 3.5-m telescope on Kitt Peak.
Spectroscopy in this field is a (small) portion of the galaxy MOS
obtained for all COS GTO fields for the purpose of determining the
relationship between gas and galaxies in the local universe (B. Keeney
et~al., in preparation).  Details of the observing strategy and
particulars, the data handling and the analysis methodology can be
found in that paper.  Briefly, for studies of the galaxy distribution
in this portion of the sky, WIYN/HYDRA MOS covers a 5 Mpc diameter
region centered on FBQS\,1010$+$3003 obtaining viable galaxy spectra
for a complete sample of galaxies down to $g=20$ or $L=0.2\,L^*$ at
the group's Hubble flow distance. These Hydra spectra were augmented
by a few single-object slit spectra obtained at the Apache Point
Observatory (APO) 3.5-m telescope with the Dual-Imaging Spectrograph
(DIS) in order to ensure that the completeness extends to at least
$g=20$ in this region. Twenty-six new galaxy redshifts were obtained
which fall within 1000 \kms\ of $z=0.113$.  The 20 galaxies which are
located at $\rho<2$ Mpc and $\pm2000$ \kms\ from the sightline are
listed in Table~2 in decreasing luminosity order.  The basic
information in Table~2 includes: (1) numerical designation used
herein; (2) the SDSS DR12 galaxy designation which includes the RA and
DEC of the galaxy (the QSO target FBQS\,1010$+$3003 has (RA,DEC) =
10h10m00.7s +30d03m22s; (3) galaxy redshift, where the redshift errors
are estimated to be $\pm30$ \kms; (4) the rest-frame $g$-band galaxy
luminosity, $L_{gal}$, in $L^*$ units; (5) the impact parameter
($\rho$) of the galaxy from the sightline in kpc; (6) the impact
parameter divided by the galaxy's virial radius ($\rho/R_{vir}$)
\citep[see][for how the virial radius is derived from the stellar
luminosity]{stocke13}; and (7) the velocity difference ($\Delta v$) of
the galaxy from the absorber in \kms.

At this point it is worthwhile to discuss briefly the limited meaning
of the virial radius (see Table 2) for individual galaxies within the
confines of spiral rich groups of galaxies.  From a theoretical
perspective, the virial radius becomes a less useful terminology and a
less-well defined term for ``sub-halos'' (individual galaxies) within
a main halo (a galaxy group).  While we recognize the ambiguities in
calculating a virial radius for a sub-halo, nevertheless, we use this
term herein to mean a characteristic radius for the CGM of an
individual galaxy which scales as the total halo mass calculated only
from the stellar luminosity; i.e., an $L^*$ galaxy has a halo mass of
$10^{12}~M_\odot$ and a virial radius of $R_{vir}=250$ kpc
\citep[see Figure 1 in][]{stocke13}.  While the $R_{vir}$ terminology
is used in Table 2 and when referring to an individual galaxy, in
Table 3 and in the context of galaxy members of a group we will use
the term $R_{sub}$ for the characteristic radius of a sub-halo.
Numerically, $R_{sub} = R_{vir}$ in this paper.

The assumption that an individual galaxy's CGM is confined largely
within the virial radius $R_{vir}$ is based upon a scrutiny of
``serendipitously'' discovered absorber-galaxy pairs carried out in
\citet{stocke13} and \citet{keeney17}. In \citet{keeney17} we showed
that the association of abosrbers with individual galaxies is quite
secure at $\rho\leq1.4\,R_{vir}$.  Absorbers at larger impact
parameters often ($\sim1/3$ of cases) have ambiguous galaxy
associations or are plausibly associated with entire small groups of
galaxies (see also Paper 2). Additionally, low-ion metal-bearing
absorbers are found at $\rho \la R_{vir}$ excepting \HI$+$\OVI-only
absorbers which are found out to impact parameters of nearly 1~Mpc
from the nearest bright galaxy \citep{stocke06, stocke13, keeney17},
many of which likely are associated with entire groups of galaxies
(Paper 2).  But there is no indication that the virial radius
($R_{vir}$, or $R_{sub}$ in the group context) is a firm boundary for
the CGM of an individual galaxy so we use it here only as an indicator
that the absorber might or might not be associated with the CGM of an
individual galaxy. In this paper, individual galaxies are assigned
virial radii and group galaxies an $R_{sub}$ based on their rest-frame
$g$-band luminosity using a hybrid method that employs a
halo-abundance matching scheme at $L\leq0.5\,L^*$ \citep{trenti10} and
assumes $M_{vir}/L_{gal} = 50~M_\odot/L_\odot$ at $L\geq0.5\,L^*$.
The halo mass and virial radius as a function of galaxy luminosity are
shown in Figure~1 of \citet{stocke13}.

Returning to the specific case of the FBQS\,1010$+$3003 sightline at
$z=0.112$-0.118, we have used the Berlind group finder process of
Paper~3 on all 20 galaxy locations and redshifts considered as
possible group members. With these new data the original Berlind group
fragments into 4-5 groups corresponding to the five brightest
galaxies.  This is because the numbers of satellite galaxies in this
region do not increase with decreasing luminosity as fast as a
standard SDSS luminosity function, decreasing the adopted linking
length relative to the galaxy separations both in redshift and on the
sky.  The friends-of-friends method of Paper~3 is more suited to a
large homogeneous sample of galaxies rather than the small region
around a single group that we have here. We therefore adopt a new
approach to establishing group membership, where we start by using
each galaxy's stellar mass to estimate a total halo mass using the
observed tight correlation (0.2 dex spread in stellar mass at a fixed
halo mass with the converse spread being somewhat larger than this)
found by \citet*{behroozi10}. This halo mass is likely overestimated
because we have ignored the scatter in the stellar-to-halo mass
relation, making the group finding err on the side of being inclusive.
The total galaxy mass then determines a virial radius, which can be
plotted on the sky to determine if these overlap.  If a satellite
galaxy is encompassed by the projection of this radius on the sky, the
velocity difference between the central and satellite galaxy is
checked to see if this velocity difference can be accomodated at high
($>90$\%) probability for a satellite bound to its primary given the
inferred halo mass of the primary.  This procedure results in three
groups and a number of single galaxies in this region. These include a
group with five satellites to the most luminous SDSS galaxy and two
other groups of two galaxies each around the 3rd and 5th brightest
galaxies.

The most abundant group has a mean velocity of $z=0.11662$ with
$\sigma_v\approx200$ \kms\ and is centered on the brightest galaxy
(\#1 at $2.2\,L^*$ and $z=0.11551$), which is nearly 1~Mpc away from
the sightline and 600 \kms\ from the absorber in velocity (see Figure
1).  The six galaxies in this group are listed in Table~3 in order of
increasing impact parameter from the AGN sightline, with the following
information by column: (1) numerical designation from Table~2; (2)
SDSS DR12 galaxy designation; (3) galaxy redshift; (4) the rest-frame
$g$-band galaxy luminosity, $L_{gal}$, in $L^*$ units; (5) the impact
parameter of the galaxy from the sightline in kpc; (6) the
characteristic radius, $R_{sub}$, determined as described above; and
(7) the stellar mass, $M_*$, derived from the galaxy's $i$-band
absolute magnitude and rest-frame $(g-i)$ color using Equation~8 of
\citet{taylor11}.  We correct to rest-frame colors and magnitudes using the
$K$-corrections of \citet{chilingarian10} and \citet{chilingarian12}.

A less conservative approach to group finding which was adopted in
Paper~2 uses the friends-of-friends approach but with considerably
larger linking lengths (5~times the virial radius for each galaxy)
than in Paper~3. Examples of groups identified by this process can be
seen in Figure 2 of Paper~2. Since these plots color-code group
members and non-members, the examples in Paper~2 show explicitly that
in almost all cases, many but not all galaxies in these regions are
linked into a single group of galaxies by this method.  Similar to the
procedure in Paper 3, once the group membership is established, all
galaxies (i.e., not just those identified as group members) within
$\pm3\sigma$ of the group redshift and $1.5\,R_{grp}$ of the group
centroid are included in the group.  All galaxies surveyed by us at
$\leq2$~Mpc and $\pm2000$ \kms\ from the absorber are included in the
group finding analysis.  This procedure is applied in a iterative
fashion until it converges which occurs quickly; i.e., usually in
$\leq3$ iterations.  Because these small groups of galaxies may not be
completely virialized (similar to the Local Group), group finding and
membership determination are not exact processes; we used both
approaches here in order to provide useful bounds on the groups that
are present in this region and their memberships. By way of an
example, if these two procedures were applied to the Local Group, the
less conservative approach of Paper~2 would identify the Milky Way and
Andromeda as being members of the same group whereas the more
restrictive approach adopted in Paper~3 would identify these two
galaxies as being in separate groups.

By this procedure a 7 member, $4.3\,L^*$ group is identified that is
almost identical to the group defined by the more conservative
approach discussed above.  Only galaxy \#9 is added and so the group
properties have not changed appreciably between these two group
finding methods; data for this galaxy have been added to Table 3 of
group members.  Only one of the three bright galaxies in the original
friends-of-friends group (\#49880) is included in this newly-defined
group.  The membership and extent on the sky for this group are shown
in Figure 3, which includes all six galaxies identified by the
previous procedure plus one other (\#9) identified by the more relaxed
procedure. In Figure~3 the color-coded symbols are the identified
group members with red and blue symbols indicating galaxies that are
redshifted and blueshifted respectively from the velocity centroid of
$z=0.11685$, which is 967 km s$^{-1}$ higher than the absorber
velocity. The estimated group barycenter on the sky is marked with a
black ``$X$'' and is 840 kpc from the sightline.  The group velocity
dispersion of $193^{+71}_{-90}$~\kms\ (90\% confidence interval) is
somewhat less than the virialized velocity dispersion of 262 \kms\
(which assumes $M/L=500$; see detailed definition and discussion in
Paper 2).  This indicates that as defined this group may not yet be
fully virialized. The group velocity centroid is $5\sigma_v$ from the
absorber velocity.  As with the smaller group identified by the more
conservative process, the group centroid impact parameter and velocity
difference from the absorber are substantial.

The values in Tables 2 and 3 for the virial radii of the galaxies in
this region use the ``hybrid'' relationship shown in Figure~1 of
\citet{stocke13}.  But converting a total group luminosity into a halo
mass and virial radius for the entire group is more problematical.
While the \citet{trenti10} and \citet{moster10} ``halo-matching''
technique seems relevant to these very large halos, they are most
appropriately applied to groups with single, dominant galaxies (i.e.,
``central'' halos), not for the loose group we have identified here.
Application of the pure halo-matching correlation shown in Figure~1 of
\citet{stocke13} to this small group yields a halo mass of
$\sim10^{14}~M_\odot$ and $R_{grp}>1$ Mpc.  But this group is not
dominated by a single, very massive galaxy (see Table 3).  The loose
group we have identified and delineated in Table 3 and Figure 3 has no
dominant central galaxy.  In this case,the conversion between stellar
mass and halo mass is likely to be a shallower function of stellar
luminosity \citep{hearin13}.  The numerical simulation results of
\citet[][see their Figure 4]{hearin13} show that there is a $\sim35$\%
difference in velocity dispersion as a function of richness for small
groups with and without a single, dominant galaxy.  In the current
case this leads to a total estimated halo mass of
$\sim4\times10^{13}~M_\odot$ and $R_{grp}\sim700$ kpc for this small
group.  This places the absorber at $1.2\,R_{grp}$.  While these
values seem appropriate for the small group we have identified,
further simulation work is required to clarify ``halo matching''
schemes for halos appropriate to the sizes of small galaxy groups.

In Figure 3 the diamond symbols are galaxies in this sky and redshift
region which are {\em not} identified as group members by this
process.  Specifically {\bf not included} in the group described above
is galaxy \#2, which is the closest galaxy to the sightline (colored
green in Figure~3).  Despite being quite close on the sky to group
members, it is not identified as being in this group mostly due to its
large $\Delta$v with respect to group galaxies.  The \OVI\ absorber
has a similarly large $\Delta$v compared to the galaxy group. Galaxy
\#2 is at a significantly lower velocity than the galaxies in the
group (see Figure 3 at top) and has a velocity very close to the \OVI\
and 2 of the \Lya\ absorbers shown in Figure~1.

We believe that these two methods span the reasonable range of
algorithms that can be used to define group membership in this region
and yet obtain a nearly identical outcome. The absorber impact
parameter and velocity difference relative to the identified group are
substantial in both cases (see Figure 3). A much closer match is made
to the one, $2.0 L^*$ nearest galaxy (\#2) than to either of these
possible group configurations.  Therefore, while possible, the
association of the warm gas absorber detected in \OVI\ with the group
listed in Table~3 is quite unlikely.

The identification of the broad \OVI\ with galaxy \#2 is bolstered by
the small velocity differences between this luminous, moderately star
forming spiral and three \Lya-only absorbers shown in Figure~1.  The
two \Lya\ absorbers close to the \OVI\ absorption are at velocities
($\Delta v = 67$ and 148 \kms) close to the velocity of galaxy \#2
($\Delta v=+51$ \kms). 
While galaxies \#4 and \#5 have $\Delta v$ values quite close to this
absorber, these two galaxies are also $>1.5$ Mpc from the sightline
and are not close to each other on the sky. Association with galaxy
\#2 seems much more plausible for this \Lya\ absorber as well.

The only absorber which can be plausibly associated with the galaxy
group we have identified is the highest redshift Ly$\alpha$ line in
Figure~1 (the final listing in Table~1) since it has a velocity quite
close ($\Delta v=26$ \kms) to the group velocity centroid.  However,
this absorber has no associated \OVI\ and is located greater than one
group virial radius away on the sky.

In summary, we suggest that the most plausible association of the
\OVI-only absorber found in the FBQS\,110$+$3003 sightlines at
$z=0.11326$ is a single luminous galaxy (\#2), located $1.1\,R_{vir}$
from the sightline. This same galaxy is also the most viable
associated galaxy for three velocity components of \Lya\ which are
close to, but not coincident with, the \OVI-only absorber. Three other
\Lya-only absorbers (whose velocity locations are shown in the
velocity histogram at the top of Figure~3) are likely absorbers in IGM
filaments in the vicinity of the galaxy distribution shown in the
bottom plot of Figure~3.


\section{Discussion and Conclusions}

The hypothesis put forward in Paper~2 is that warm \OVI$+$\Lya\
absorbers are associated with small galaxy groups and are detections
of interface gas between a hotter intra-group medium and the cool,
photoionized clouds amply detected in even modest S/N surveys using
HST/COS \citep{tumlinson11, prochaska11, stocke13, werk13, keeney17}.
The methodology followed in Paper~2 was to investigate the galaxies
associated with the lowest redshift warm absorbers found in Paper 1;
i.e., absorbers were discovered before the galaxy environments were
investigated.  This process led to some uncertainty about the
conclusion that warm absorbers are associated with galaxy groups
rather than individual galaxies because both galaxy groups and
luminous galaxies are comparably abundant in the low-$z$ universe.

In a continuing investigation of this hypothesis, the converse
approach is adopted here whereby a homogeneous sample of galaxy groups
selected from the SDSS spectroscopic survey by the method of Paper 3
\citep[see also the catalogue of][] {berlind08} is cross-correlated with
bright AGN targets that were already observed by HST/COS using the G130M
grating.  While six matches were found, only one previously-observed
FUV AGN spectrum has sufficient S/N to allow a sensitive search for
broad, shallow \OVI\ and \Lya\ common to these warm absorbers.  The
COS/G130M spectrum of FBQS\,1010$+$3003 shows a broad, shallow \OVI\
absorption feature at $z=0.11326$, quite close ($\Delta v=387$ \kms)
to the redshift of Berlind group \#49980 at $z=0.11455$ (Figure~1).
Only the \OVI\ 1032 \AA\ line of the doublet is detected without doubt
but there is little ambiguity in the identification because this \OVI\
-only absorption occurs blueward of the \Lya\ rest wavelength.

The observed $b$-value and rather symmetrical \OVI\ 1032\AA\ profile
suggest gas in CIE at a temperature of $5.5<\log T(K)<6.0$ although
the constraints on this conclusion are rather loose.  Higher
metallicities ($[Z]>-0.5$) are also favored but are not required
(Figure~2).  The absence of both the weaker \OVI\ doublet line and an
associated \Lya\ absorption line make the analysis of this absorber
inconclusive. When there are solid detections of the broad \OVI\
doublet and Ly$\alpha$, the $b$-values for both species allow a
determination of both the thermal and also the turbulent broadening of
the gas (Paper 1).  Higher S/N FUV spectra than the one analyzed here
are required to investigate the warm absorber population further.
$S/N>20$ G130M spectra are being obtained of 10 UV bright targets
behind 12 $z=0.1-0.2$ galaxy groups in Cycle 23.

A moderately deep (complete to $g=20$) redshift survey was conducted
on the FBQS\,1010$+$3003 field using WIYN/Hydra and the Dual Imaging
Spectrograph at the APO 3.5 meter.  While the original Paper 3
analysis of this field using the SDSS spectroscopic survey identified
a galaxy group consisting of three luminous galaxies, a new
friends-of-friends type analysis using the much deeper
($L_{limiting}=0.2\,L^*$) WIYN/HYDRA survey also found only one galaxy
group of six or seven members in this region plus a number of more
isolated galaxies.  A strict application of the original Paper 3
approach ``fragments'' this group into a number of isolated, single
galaxies, but a related, and still conservative approach, found a
group of six galaxies closely associated with the brightest galaxy in
this region at $z=0.11662$.  The remaining galaxies were found to be
either isolated galaxies or members of groups of two members only. A
less conservative friends-of-friends approach using larger linking
lengths also identified a single galaxy group of seven members, the
six previously-identified members plus only one additional, lower
luminosity galaxy.  Because the membership of these potential groups
so largely overlaps, they have similar mean redshifts which are
several hundred \kms\ greater than the \OVI\ absorber redshift and
centroid locations on the sky which are greater than 500 kpc from the
sightline. However, the velocity distribution of this larger group
does encompass the velocities of the three \Lya\ absorbers also
detected in the FBQS\,1010$+$3003 COS spectrum and there are near
coincidences in velocities between these \Lya\ absorbers and
individual bright galaxies in this group.  We identify these \Lya\
absorbers as cool gas in PIE associated with the individual CGM of
these galaxies.

The association between the warm gas detected in \OVI\ and the small
foreground group is {\bf not} clearly made due to both a large impact
parameter and a large $\Delta$v. However, the second brightest galaxy
in this region, a $2\,L^*$ spiral, is only 254 kpc ($1.1~R_{vir}$) from
the sightline and has a recession velocity which differs from the
\OVI\ absorber velocity by only $\sim50$ \kms.  Based on these close
correspondences, the warm gas detected in \OVI\ is much more likely
associated with this individual spiral galaxy than the entire galaxy
group.

While the discovery of a broad, shallow \OVI\ absorber that is
demonstrably ``warm'' in the only viable archival COS spectrum was
exciting, it has provided a cautionary tale for the investigation of
warm gas in galaxy groups.  Despite $S/N=15$ per resel at the \OVI\
doublet and $S/N=20$ at the location of \Lya, the spectrum was of
insufficient quality to fully characterize the absorber due to a
marginal detection of \OVI\ 1038~\AA.  Further, the close proximity
(impact parameter $\approx R_{vir}$) of the sightline to a single
bright galaxy made the interpretation that the warm gas was associated
with the entire galaxy group very unlikely.  Our Cycle 23 program to
use COS spectra to probe low-$z$ galaxy groups from an update to the
catalog of Paper 3 (GO program \#14277; JTS, PI) needs to avoid these
possible pitfalls.  Higher $S/N$ than found in the FBQS\,1010$+$3003
spectrum has been proposed for this program.  The impact parameters
from the chosen AGN sightlines to nearby galaxies have been required
to be $>1.5 R_{vir}$.

The HST/COS program designed to probe spiral-rich galaxy groups has
the potential to discover warm gas associated with galaxy groups,
measure its covering factor inside the group virial radius and make a
rough estimate of the mass in warm gas in these systems. As support a
deep galaxy redshift survey complete to $g\approx 21.5$ is underway at
the MMT Observatory using the Hectospec multi-object spectrograph to
characterize these groups in membership, velocity dispersion, and
group centroids in velocity and on the plane of the sky. While a
virialized intra-group medium for these groups is predicted to be too
hot to detect by the broad \OVI\ and broad \Lya\ method
\citep{mulchaey00}, detecting the warm interface gas will assist in
estimating the properties of this hotter gas.  In this paper we have
highlighted some of the difficulties in testing the hypothesis of
Paper~2 that the warm and hot gas reservoirs in galaxy groups may
contain the bulk of the baryons still ``missing'' from spiral galaxy
halos.

Facilities: HST/COS, WIYN, APO 3.5m

\section{Acknowledgments}

This research was supported by NASA/HST Guest Observing grant \#14277 
(PI: JTS). BDO's contributions were supported by a NASA/HST Archive/Theory grant
\#14308.

\end{document}